\documentclass[twocolumn,secnumarabic,amssymb, nobibnotes, aps, prl, superscriptaddress]{revtex4-1}
\usepackage{todonotes}

\usepackage{natbib}
\usepackage{graphicx}
\usepackage[inkscapeformat=png]{svg}
\usepackage{amsmath}
\usepackage{empheq}

\usepackage{tabularx}
\usepackage{lineno}

\usepackage{lineno}

\begin{document}

\title{Phase noise properties of supercontinuum generation in all-normal dispersion fibers}

\author{Matis Marcadier}
\email{matis.marcadier@univ-cotedazur.fr}
\affiliation{Universit\'e C\^{o}te d'Azur, CNRS, Institut de Physique de Nice (INPHYNI), UMR 7010, 17 rue Julien Laupr\^{e}tre, 06 200 Nice, France}
\affiliation{Fastlite, 165 route des Cistes, 06600 Antibes, France}

\author{Nicolas Forget}
\affiliation{Universit\'e C\^{o}te d'Azur, CNRS, Institut de Physique de Nice (INPHYNI), UMR 7010, 17 rue Julien Laupr\^{e}tre, 06 200 Nice, France}

\author{Yoann Pertot}
\affiliation{Fastlite, 165 route des Cistes, 06600 Antibes, France}

\author{Aur\'elie Jullien}
\affiliation{Universit\'e C\^{o}te d'Azur, CNRS, Institut de Physique de Nice (INPHYNI), UMR 7010, 17 rue Julien Laupr\^{e}tre, 06 200 Nice, France}

\begin{abstract}
The spectral coherence properties of supercontinuum generation in polarization-maintaining all-normal dispersion fibers are investigated. Stochastic phase noise induced by energy fluctuations, along with spectrally-resolved intensity-to-phase transfer coefficients, are quantitatively analyzed, confirming the high coherence of the generated supercontinuum. Our results show that the nonlinear process is fundamentally deterministic with ultra-low spectral phase noise, yet exhibits significant intensity-to-phase coupling.
\end{abstract}


\maketitle

Supercontinuum generation (SCG) in nonlinear media is a highly effective technique for spectrally broadening femtosecond pulses, often extending the spectrum by up to an octave. As such, SCG plays a central role in systems that deliver few-cycle pulses and/or shot-to-shot carrier-envelope phase (CEP) stabilized pulses, serving both pulse generation and metrology applications.
Driven primarily by self-phase modulation, SCG can occur in a variety of media, ranging from crystals and gases to hollow-core fibers and photonic crystal fibers, with either normal or anomalous dispersion. A fundamental characteristic of SCG is the pulse-to-pulse phase relationship between the spectral components of the continuum, commonly known as intrapulse coherence \cite{PhysRevLett.119.123901}. This coherence is essential for precise pulse compression and coherent combining \cite{Newton:25}. Furthermore, its degradation can be a significant source of CEP noise.

In prior work, we introduced a modified Bellini-Hänsch interferometer, termed a double white-light interferometer, to experimentally probe the intrapulse coherence of supercontinuum generation (SCG) in bulk materials, where the process is driven by filamentation (filamentation-SCG) \cite{Bellini:2000uk,Maingot:22,Maingot:2024aa}. The stability of the spectral phase can be systematically characterized based on the parameters of the laser system used to initiate filamentation-SCG, including its central wavelength, pulse duration, pulse energy, and average power. Two experimental scenarios emerge: either applying controlled perturbations to both SCG stages simultaneously or introducing perturbations selectively into a single arm (test arm). The latter approach enables precise quantification of phase fluctuations relative to the reference arm.
These differential perturbations include variations in pulse energy, the position of the nonlinear crystal relative to the beam waist and the spatial phase of the pump beam. Our experiments in YAG bulk crystals revealed three key findings: (i) the coherence of the filamentation-SCG process is highly sensitive to the pump laser parameters; (ii) high spectral coherence can be preserved under optimized conditions; and (iii) a stochastic noise floor remains inherent to the system. Additionally, we measured frequency-resolved intensity-to-phase coupling coefficients, which exhibited a strong dependence on the filamentation-SCG operating range.

Building on previous work, we present here a quantitative analysis of the phase noise properties of supercontinuum generation in polarization-maintaining (PM) all-normal dispersion (ANDi)  optical fibers. These fibers have recently attracted considerable interest for SCG applications due to their ability to produce highly coherent, low-noise spectra at low energy levels \cite{Genier:19, Genier:20, Sylvestre:21, Klimczak:2016aa}. Their spectral broadening mechanism, dominated by self-phase modulation without ionization, enables the production of intense, uniform, and stable supercontinua with minimal relative intensity noise (RIN). Consequently, they are anticipated to deliver superior spectral coherence in the visible range compared to filamentation-SCG \cite{Liao:2020aa,Mei:20}.
In this paper, we then present the experimental characterization of intrapulse coherence of SCG in a commercial PM-ANDi fiber. The stochastic phase noise is measured continuously over the full spectral bandwidth under various conditions of seed energy and intensity-to-phase coefficients are assessed.

\begin{figure}[ht]
\centering
\includegraphics[width=1\linewidth]{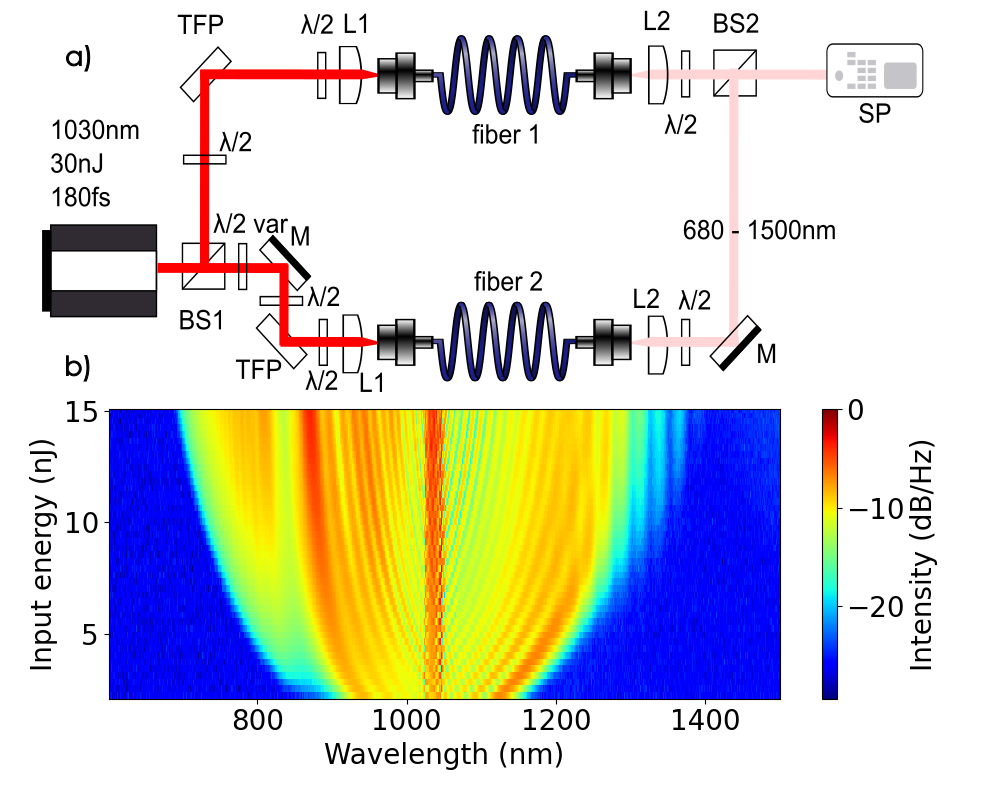}
\caption{(a) Experimental setup. BS1 : 1030\,nm 50/50 beamsplitter, BS2: 600-1500\,nm 50/50 beamsplitter, $\lambda$/2: half wave plate, TFP: thin film polarizer, L1: f = 3.14\,mm, L2: achromat f = 8\,mm, F 1 F2 : $~$12\,cm NL-PM-1050-NEG NKT Photonics fiber, SP : Spectrometer. (b) Spectral intensity profile in log-scale as a function of laser input energy (reference arm, slow axis).}
\label{fig:setup}
\end{figure}

The tested PM-ANDi fibers are commercially available from NKT Photonics under the reference NL-PM-1050-NEG. These fibers feature a parabolic all-normal dispersion profile with a minimum at 1050\,nm (maximum group velocity dispersion is -8$\pm$6\,ps/nm/km, with a weak chirp coefficient of $\simeq 5\,\text{fs}^2\text{/mm}$). A weak birefringence induced by stress-rods ($\Delta n = 1.7\times 10^{-4}$) ensures polarization maintaining properties, enhancing coherence properties \cite{Liu:2015,Gonzalo:2018aa}. In the following, the fiber eigenaxes are referred to as the slow and fast axis. 
The nonlinear index is estimated to $n_2 \simeq 2.5 \times 10^{-16} \text{cm}^2\text{/W}$ \cite{Kuliesaits:24}. 
The pump laser is a regenerative CPA system (Pharos, Light Conversion) delivering pulses at $\simeq$1035\,nm with a pulse duration of $\simeq$180\,fs FWHM. For these experiments, the repetition rate is set at 1 kHz and $\simeq$100\,nJ are sampled from the available 1\,mJ of pulse energy, with an energy stability better than 0.5\%\,rms. 
The interferometer is of Mach-Zehnder type (Fig. \ref{fig:setup}a) with each arm including a half-waveplate and a polarizer to adjust the pulse energy independently. An additional half-waveplate controls the polarization at the fiber input. A pair of aspheric lenses with a focal length of 3.14\,mm is used to couple the light into the 2.4\,$\mu$m fiber core and collimate the spectrally-broadened beams. Both fibers are 12-cm-long. The coupling efficiency is $\simeq \text{50}\%$ and an achromatic half-waveplate sets the output beam polarization to horizontal. 

The spectral intensity profile at the output of one fiber, as a function of input pulse energy, is shown in Fig. \ref{fig:setup}b. The laser is polarized along the slow axis of the fiber and the laser compression is optimized so as to get the broadest spectrum. Spectra recorded with a broadband scanning spectrometer (APE Wavescan) display smooth spectral broadening with SPM features emerging for central wavelengths. For a coupled seed energy of $\simeq$7.5\,nJ (15\,nJ at the input of the fiber), the input peak intensity of $\simeq 1.3\,\text{TW/cm}^2$ (peak power of 38\,kW), and the SC extends from roughly 700\,nm to 1400\,nm.

\begin{figure}[ht]
\centering
\includegraphics[width=1\linewidth]{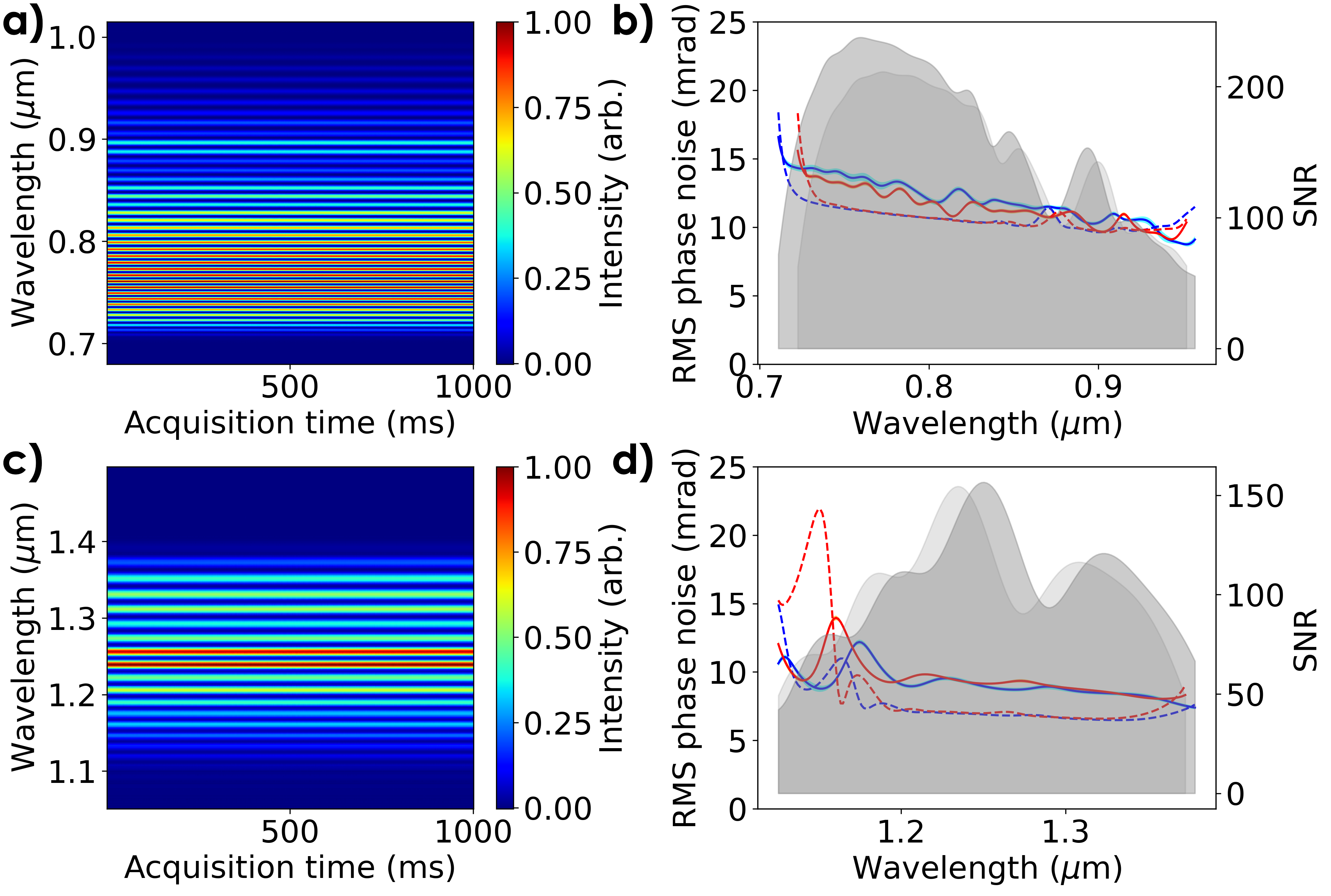}
\caption{(a,c) Spectrograms for the short- and long-wavelength spectral ranges, obtained from 1000 single-shot measurements. (b,d) Phase standard deviation ($\sigma$, solid lines) as a function of wavelength (left y-scale), measured along the two fiber's eigen-axis (blue : slow axis, red : fast axis). The dashed lines show the noise model $\sigma_\text{meas}$, according to Eq. \ref{eq:3} (blue : slow axis, red : fast axis). The shaded area indicate the experimental SNR for both axis (right y-scale).}
\label{fig:sigma}
\end{figure}

The interferometer consists of two arms: one serving as the test arm, where pulse energy is the adjustable parameter, and the other as the reference arm. A reflective delay line precisely controls the relative group delay between the two arms. After recombination, the beams are directed into two array-based spectrometers for spectral analysis: a silicon spectrometer covering the 450–1100\,nm range (0.26\,nm resolution, 1.1\,ms integration time) and an InGaAs spectrometer spanning 900–1700\,nm (2.8\,nm resolution, 1\,ms integration time).
When the relative group delay is a few hundred femtoseconds, interference fringes are observed across the entire spectral range, as shown in Fig.\,\ref{fig:sigma}. To evaluate relative phase stability, 1000 consecutive single-shot interference spectra are recorded and analyzed using Fourier-transform spectral interferometry (FTSI) \cite{Lepetit:95,Borzsonyi2013What-We-Can-Lea}. The standard deviation of the phase ($\sigma$) at each wavelength serves as a measure of the frequency-resolved stochastic phase noise. A 5\,Hz high-pass filter is applied to eliminate slow phase drifts and wavelengths where the spectrometer's noise matches the detected signal level are excluded.
Whether without fibers or with fibers seeded at very low energy (1\,nJ), the  typical measured standard deviation remains 8$\pm2$\,mrad at 1035\,nm (which is equivalent to fluctuations of the relative optical path of $\simeq$1.5\,nm rms), establishing the measurement noise floor. For the following, the pulse energy in the reference arm is set to 15\,nJ.

Figures \ref{fig:sigma}(a,c) show the spectrogram registered for both wavelength ranges when the interferometer is balanced, e.g. seeded with 15\,nJ in both arms. The retrieved phase stability is shown in Fig. \ref{fig:sigma}(b,d). The phase noise is extremely low (between 10 and 15\,mrad, with $\simeq 3\%$ experimental deviation evaluated from several successive measurements) and mostly constant across the full measurement range. 

The experimental results are compared with an estimate of the phase noise introduced by the measurement itself, denoted $\sigma_\text{meas}$. This quantity arises from the combination of two independent stochastic contributions. The first source, $\sigma_\text{interf}$, corresponds to environmental variations of the optical path difference in the interferometer, measured at 1035\,nm ($\simeq$1.5\,nm rms) and extrapolated across the entire spectrum. The second source, $\sigma_\phi$, originates from the FTSI algorithm, which maps the spectrometer measurement noise (mainly dark and shot noise \cite{konnik2014high}) into phase noise. 
The FTSI algorithm consists of four steps: (1) discrete Fourier transform (DFT), (2) multiplication by a gate function of width $\Delta\tau$ centered at delay $\tau$, (3) inverse DFT, and (4) computation of the argument (phase) of the resulting complex values. Steps 1–3 can be viewed as a convolution of the input spectrum with a complex-valued gate function $G(\omega_n)$. 
Let $S(\omega_n) + \delta S_k(\omega_n) = S_n + \delta S_{n,k}$ represent the $k^\text{th}$ interferogram, where $S_n = I_n[1 + V \cos(\omega_n\tau + \phi_n)]$ is the noiseless interferogram, $\phi_n$ is the noiseless relative phase between the two SCG, and $\delta S_{n,k}$ are independent normal variables with null average. Here, $\omega_n = n\delta\omega$ is the discrete angular frequency, with $\delta\omega$ being the spectral resolution.
The convolution of $S_n$ with $G_n$ gives:
\begin{equation}\label{eq:1}
    S_{n,\text{conv}} = \frac{I_n}{4}\exp\left[i\omega_n\tau + \phi_n\right]
\end{equation}
Since $\delta S_{n,k}$ are real-valued Gaussian processes and $G_n$ is complex, the convolution $\delta S_{n,k} * G_n$ is a complex-valued Gaussian process. Assuming the $\delta S_{n,k}$ are independent with $\sigma_n$ standard deviation, $\delta S_{n,k} * G_n$ has circular symmetry, with a variance $\sigma^2_{n,\text{conv}}$ for both the real and imaginary parts.
Adding this complex Gaussian noise to $S_{n,\text{conv}}$ introduces uncertainty in the retrieved phase. In the high signal-to-noise ratio (SNR) limit, the phase standard deviation is:
\begin{equation}\label{eq:2}
    \sigma_{n,\phi} = \frac{\sigma_{n,\text{conv}}}{<|S_{n,\text{conv}}|>} = \frac{1}{\text{SNR}_{n,\text{conv}}}
\end{equation}

Finally, the expected phase noise at $\omega_n$, including both interferometer and FTSI contributions, is:
\begin{equation}\label{eq:3}
     \sigma_{n, \text{meas}} = \sqrt{\sigma^2_{n, \text{interf}} +\sigma_{n, \phi}^2}  = \sqrt{\sigma^2_{n,\text{interf}} + 1/\text{SNR}^2_{n,\text{conv}}}
\end{equation}

Figures \ref{fig:sigma}(b,d) compare the experimentally retrieved phase stability with the predictions of the noise model ($\sigma_\text{meas}$, Eq.\ref{eq:3}). The two data sets are in very good agreement, with the measured phase noise slightly exceeding $\sigma_\text{meas}$. The wavelength-dependent variations of the phase noise closely follow those of the SNR, indicating that the measurement is primarily limited either by mechanical noise at high SNR or by the finite SNR at lower values. Overall, the phase stability is excellent along both axes of the PM-ANDi fibers, and the process is highly deterministic.

\begin{figure}[ht]
\centering
\includegraphics[width=1\linewidth]{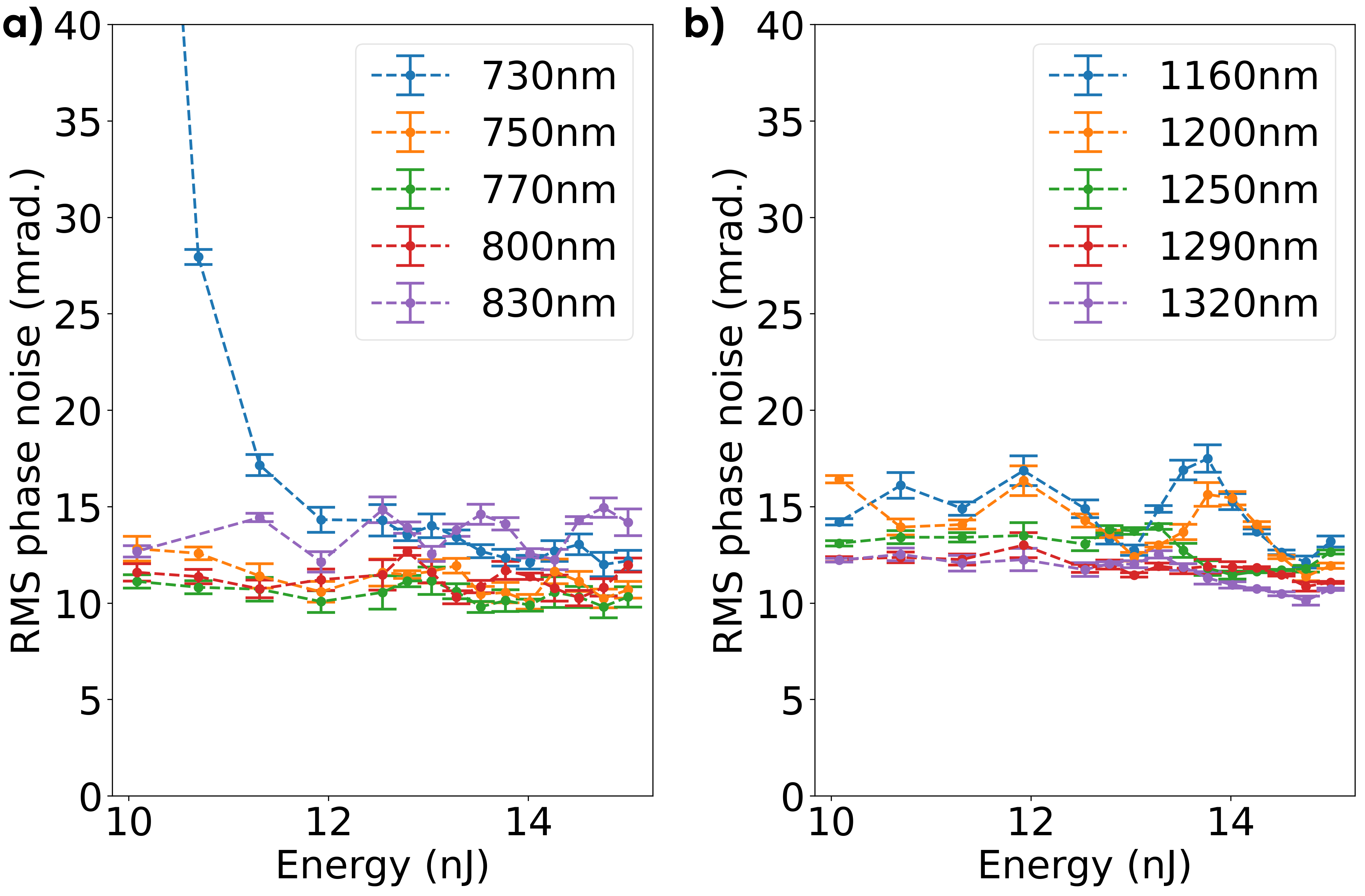}
\caption{Phase standard deviation as a function of seed pulse energy of the test arm, in the short (a) and long (b) wavelength ranges. The energy in the reference arm is 15 nJ.}
\label{fig:Escan}
\end{figure}

To study the dependence of the phase noise with the input peak power, the energy in the test arm is varied from 10 to 15\,nJ (steps of 0.25\,nJ) (Fig. \ref{fig:Escan}). Both short and long wavelength ranges display a highly stable behavior with a rms phase noise slightly above 10\,mrad, whatever the energy and the selected wavelength. Only the extreme wavelengths (730 nm and below) exhibit a phase noise dependence on pulse energy, primarily due to resulting lower signal-to-noise ratio on spectral edges.
\begin{figure}[ht]
\centering
\includegraphics[width=0.8\linewidth]{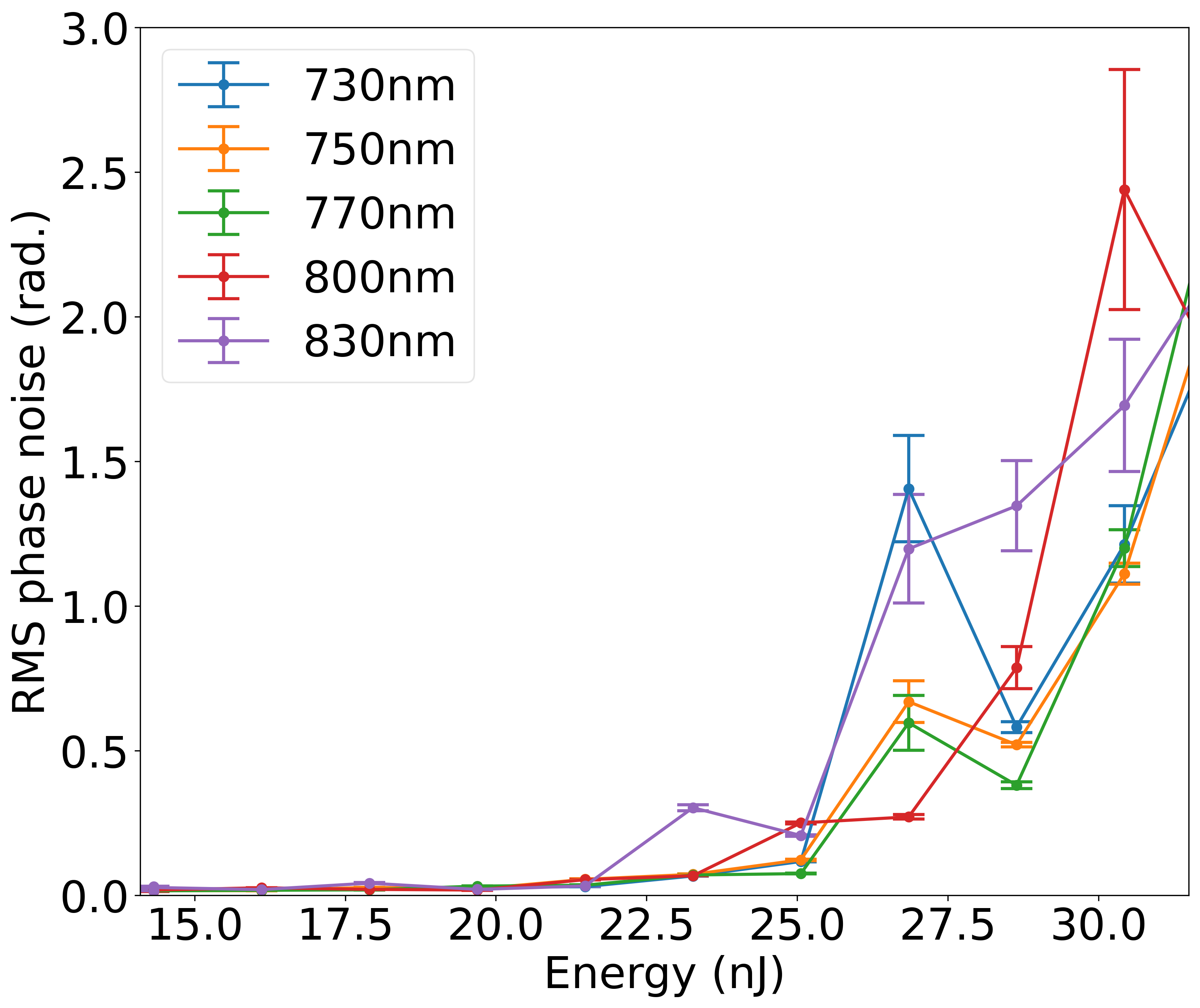}
\caption{Phase standard deviation in the short wavelength range at high input energy. The seed energy is increased simultaneously in both arms. }
\label{fig:highE}
\end{figure}
These results demonstrate the high level of intra-pulse coherence of SCG in PM-ANDi fibers. To identify the energy threshold at which coherence degrades, the laser energy is increased in both arms up to 30\,nJ. The resulting phase noise in the visible spectral range is shown in Figure \ref{fig:highE}. The excellent noise performance is maintained up to a seed energy of 20\,nJ. At 25\,nJ, the phase noise increases by one order of magnitude, and at 30\,nJ, coherence is effectively lost, which can be attributed to the occurrence of other non linear effects, such as ionization.

\begin{figure}[ht]
\centering
\includegraphics[width=1\linewidth]{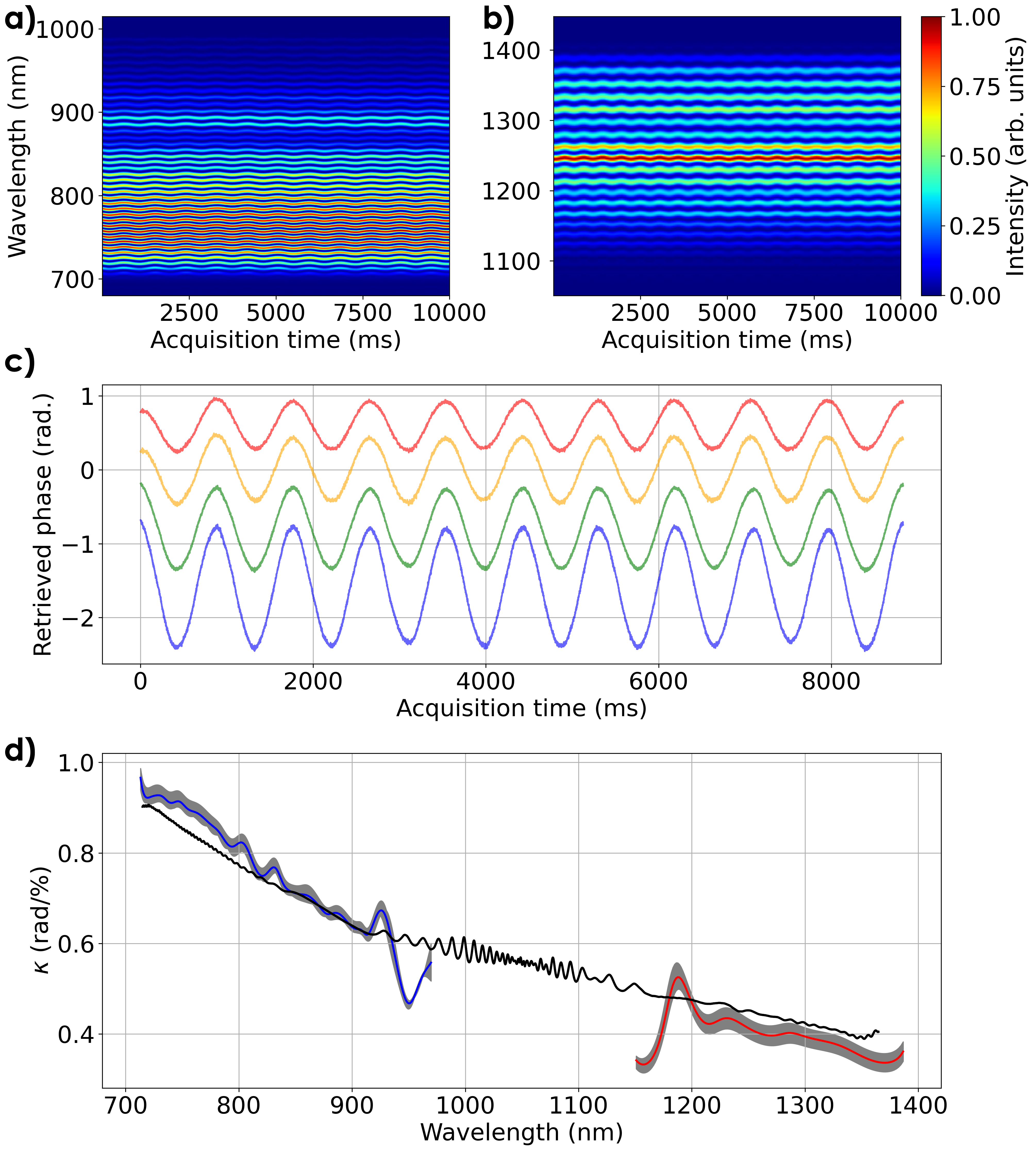}
\caption{(a,b) Spectrograms measured over 10000 single shot measurements with time-varying input energy in one arm. The fiber slow axis is selected in the two arms. (c) Retrieved phase for selected wavelengths (1300\,nm, 1200\,nm, 890\,nm, 710\,nm - from top to bottom). The curves are artificially offset. (d) Experimental intensity-to-phase coupling coefficient  $\kappa(\lambda)$ (rad$/\%$) as a function of wavelength (blue for the short wavelength range and red for the long-wavelength range). The gray-shaded region indicates the experimental deviation, evaluated from three successive measurements. The black solid line is the result of the numerical simulation.}
\label{fig:kappa}
\end{figure}

Finally, the impact of energy fluctuations on the SCG process is characterized by the intensity-to-phase coupling coefficient, $\kappa(\lambda)$, which quantifies the phase shift induced by a 1$\%$ change in intensity. 
To do so, a variable half-waveplate is implemented in the test arm, to introduce a sinusoidal energy modulation of $1.6\%$. 
Figures \ref{fig:kappa}(a,b) display the measured spectrograms (from 10,000 acquisition scans). The imposed energy modulation is indeed transferred into a phase modulation, well above the phase noise level, as shown in Fig. \ref{fig:kappa}(c). The phase-shift enables the calculation of $\kappa(\lambda)$, plotted in Fig. \ref{fig:kappa}(d) over the full spectrum. $\kappa(\lambda)$ is not measured near the fundamental wavelength of 1030\,nm due to insufficient signal, which results from spectral filtering in both spectrometers. The measurements are validated against home-made numerical simulations of the SCG by solving the generalized nonlinear Schrödinger equation, which includes both self-steepening and Raman effects. The simulations incorporate the experimental pulse intensity (spectral amplitude and phase) as well as the measured fiber dispersion and are in good agreement with experimental data. A pulse energy of 7.3\,nJ in the fiber is considered. The small discrepancy between experiment and simulation likely arises from uncertainties in the input pulse’s spectral phase and amplitude.
Across both spectral regions, $\kappa(\lambda)$ exhibits an approximate inverse-wavelength dependence, consistent with a deterministic mechanism governed by self-phase modulation, as expected in a short fiber.
$\kappa$ measured at 800 nm is $\simeq 0.8\,\text{rad/}\%$. This value is consistent with a simple estimate of the nonlinear phase shift, which is on the order of a hundred radians at 800\,nm, assuming negligible fiber dispersion. This means that pump energy stability around 0.12\% is required to achieve 100 mrad phase precision. 

These results, together with our previous work \cite{Maingot:22,Maingot:2024aa}, allow us to outline a comparison of the spectral phase stability properties between SCG by filamentation in crystals and SCG dominated by self-phase modulation in fibers. It should be recalled that the energy range differs by three orders of magnitude (µJ for crystals, nJ for fibers). The shot-to-shot reproducibility of the spectral phase in SCG by filamentation strongly depends on the pump laser quality and the chosen experimental configuration. In particular, it is difficult to achieve excellent stability simultaneously in both the visible and infrared ranges. However, under specific conditions (generating the filament at the threshold of double filamentation), the intensity-to-phase transfer coefficients can be canceled across the entire spectrum.
By contrast, SCG in ANDi fibers exhibits excellent shot-to-shot phase stability across the whole generated spectrum, regardless of energy (up to 20\,nJ), which is remarkable. One may consider that residual imperfections of the pump laser are smoothed out by modal filtering, mainly affecting the coupling into the fiber. Thus, the value of the intensity-to-phase transfer coefficient becomes critical. We have shown that this coefficient is deterministic, linked to the amount of spectral phase accumulated during propagation in the fiber. It can therefore be fully anticipated, and could potentially be modified by adjusting the fiber’s dispersion parameters (dispersion shape, zero-dispersion wavelength) in order to alter the balance between dispersion and nonlinear processes.

To conclude, these experimental results quantify the phase noise of supercontinuum generated in PM-ANDi fibers and establish a very low phase-noise floor level of 10-15\,mrad rms, depending on the wavelength, in the 5\,Hz and 1\,kHz bandwidth.  The phase noise is shown to be mostly limited by the measurement noise. This excellent intra-pulse coherence is robust to seed energy changes, up to 20\,nJ (1.7\,TW/cm$^2$ peak intensity, 49\,kW peak power).  PM-ANDi fibers thus exhibit a deterministic behavior but present a significant intensity-to-phase coupling, as expected from a process mainly driven by self-phase modulation and as reproduced by numerical simulations. These results are relevant for systems based on intra-pulse frequency mixing and  provide quantitative estimates to guide the development of future ultra-stable and ultra-broadband sources. These findings are also pertinent for carrier-envelope phase (CEP) metrology using f-to-2f interferometry, as well as for other spectral broadening techniques employed in the generation of few-cycle laser pulses \cite{Nagy:2020aa,Hanna:2021vk}.

	\section{Funding} European Union’s Horizon research and innovation program under grant agreement No. 101135904 (VISUAL project). Agence Nationale de la Recherche (ANR-19-CE30-0006-01); Association Nationale de la Recherche et de la Technologie (2023/1233).
	
	
	\section{Disclosures} The authors declare no conflicts of interest.
	
	\section{Data Availability Statement} Data underlying the results presented in this paper are not publicly available at this time but may be obtained from the authors upon reasonable request.



\bibliography{/Users/Aurele/Documents/Recherche/Bibliographie/bibtex}



\end{document}